\documentclass[runningheads]{llncs}
\usepackage[T1]{fontenc}
\usepackage{graphicx}
\usepackage{latexsym}
\usepackage{graphicx}
\usepackage{mathptmx}
\usepackage{amsmath}
\usepackage{amsfonts}
\usepackage{amssymb}
\usepackage{amsbsy}
\usepackage{booktabs}
\usepackage{cite}

\usepackage[pdftex,colorlinks=true,urlcolor=blue,citecolor=black,anchorcolor=black,linkcolor=black]{hyperref}

\begin{document}

\title{Simulation of Stance Perturbations}

\author{Peter Carragher\orcidID{0000-0003-4053-1001} \and
Lynnette Hui Xian Ng\orcidID{0000-0002-2740-7818} \and
Kathleen M. Carley\orcidID{0000-0002-6356-0238}}

\authorrunning{Carragher et al.}
%
\institute{Carnegie Mellon University, Pittsburgh, USA \\
\email{\{pcarragh, huixiann, kathleen.carley\}@andrew.cmu.edu}}

\maketitle              
\begin{abstract}
In this work, we analyze the circumstances under which social influence operations are likely to succeed. These circumstances include the selection of Confederate agents to execute intentional perturbations and the selection of Perturbation strategies. We use Agent-Based Modelling (ABM) as a simulation technique to observe the effect of intentional stance perturbations on scale-free networks. We develop a co-evolutionary social influence model to interrogate the tradeoff between perturbing stance and maintaining influence when these variables are linked through homophily. In our experiments, we observe that stances in a network will converge in sufficient simulation timesteps, influential agents are the best Confederates and the optimal Perturbation strategy involves the cascade of local ego networks. Finally, our experimental results support the theory of tipping points and are in line with empirical findings suggesting that 20-25\% of agents need to be Confederates before a change in consensus can be achieved.

\end{abstract}

\section{Background}
ABM has a rich history in social network analysis \cite{friedkin, construct, hopfield_sim, disinfo_sim, manipulative_bots, abm_sna_review}. Will et. al. \cite{abm_sna_review} categorized such studies into those that investigate endogenously emerging networks, exogenously imposed networks and co-evolutionary networks. Endogenous studies look at how social network structure evolves as a function of the set of agent states. Conversely, exogenous studies keep the network structure constant and model changes in agent states based on this structure. The co-evolutionary approach is a hybrid of both and models the interplay between agent states and network structure. Despite being a closer fit to genuine SI processes, this approach is relatively understudied \cite{abm_sna_review}. Differing timescales between endogenous \& exogenous effects complicate the matter.

Ng et. al. \cite{stance_flip} demonstrated that endogenous and exogenous features are equally important in predicting pro / anti-vaccine stance flips on Twitter. The SI model used to make this prediction is an exogenous one based on Friedkin's foundational social influence (SI) model \cite{friedkin}, where the influence weight matrix $W$ is static and pre-calculated from self-reports. In contrast Macy et. al. \cite{hopfield_sim} develop the Hopfield model, a co-evolutionary approach for the simulation of stance change. It has been shown that both exogenous \cite{disinfo_sim} and co-evolutionary models can converge to a hyper-polarized state \cite{hopfield_sim}.

ABM has also been used to highlight the vulnerability of the SI model to manipulative actors \. Ross et. al. \cite{manipulative_bots} find that as little as 2-4\% well-positioned bots are capable of tipping the majority opinion. Of note is the variance in estimations of tipping points, with Centola et. al. \cite{tipping_points} showing that 25\% of users must commit to certain language use before it gains traction. This motivates the investigation of a wide variety of environmental conditions in the simulation of intentional stance perturbation.  

\section{Simulation Objectives}
In our simulation, we wish to observe the effect of intentional stance perturbations on the overall stances in the network. The process of a group of Confederate agents perturbing a social network simulates the real-world scenario of an influence operation, wherein disingenuous operatives strategically promote subversive and provocative content with the intention of catalyzing a shift in or preventing the formation of consensus.

\subsection{Research Questions}

In this study, our objective is to explore deliberate perturbations' impact on influence networks and their effect on individual stances. Using a co-evolutionary SI model that simulates how stance changes with respect to interpersonal influence, we seek to understand strategies that maintain influence while perturbing the network. 

\textbf{R1} How can Confederates manipulate stances effectively without losing influence?

\textbf{R2} Considering factors such as susceptibility, network structure, and stance, which agents make for effective Confederates?

Finally, we investigate the potential for intentional perturbations to disrupt the prevailing consensus, shedding light on opinion dynamics within the influence network.

\textbf{R2} To what extent can intentional perturbations change an established consensus?

By addressing these research questions, we will gain a deeper understanding of intentional perturbations in influence networks, their impact on stances, and strategies to optimize their effectiveness. 



\subsection{Contributions}
By adopting a co-evolutionary SI model, we explore the tradeoff between influence and stance within social networks. Our simulations capture the intricate relationship between the emergence of the influence network and its impact on individual stances.

Secondly, we propose an evaluation criterion based on consensus, which serves as a measure to assess intentional stance perturbation. By analyzing changes in consensus over time, we can gauge the success of perturbations by the magnitude of stance change. Furthermore, our findings reveal the efficacy of targeted nudging strategies in perturbing stances. Well-targeted nudges yield the largest stance perturbation, underscoring the importance of utilizing the structure of the influence network.

Lastly, we find support for empirical findings on the theory of tipping points that suggest 20-25\% of agents need to be Confederates to bring about a new consensus \cite{tipping_points}. These contributions significantly enhance our understanding of influence dynamics, intentional stance perturbation, and consensus formation within social networks.



\setlength{\tabcolsep}{12pt}
\begin{table*}[!ht]
\centering
\caption{Definitions of Terminology used in this study}
\begin{tabular}{ll}
\toprule
\textbf{Terminology} & \textbf{Definition} \\  
\midrule
Stance & Opinion on a topic using a scale of two extremes, -1.0 and 1.0\\  
Susceptibility & How vulnerable an agent is to the changes of stance / influence \\  
Influence & How much impact / sway an agent has towards its neighbors \\  
Confederates & The group of agents that are perturbing the network \\  
Perturbation & The ``nudging" of stances of other agents in the network \\  
Tipping point & The timestep of the simulation where majority stance changes \\  
\bottomrule
\end{tabular}
\label{tab:terminology_def}
\end{table*}

\begin{table*}[!ht]
\centering
\caption{Definitions of Symbols used in this study}
\begin{tabular}{cl}
\toprule
\textbf{Symbol} & \textbf{Definition} \\ 
\bottomrule
$i,j$ & identification of agents \\ 
$y(t)$ & stance at timestep $t$ \\  
$W$ & influence weight matrix, row-normalized \\ 
$A$ & diagonal matrix of actor susceptibilities to influence \\  
$\alpha$ & stance update rate \\  
$\lambda$ & influence update rate \\  
$\mu_y$ & average stance at time $t$ over $N$ non-Confederate agents\\ 
$\theta$ & threshold of Confederate influence \\ 
$k$ & node at timestep $t$ with maximum global influence \\  
$w_i^g$ & Confederate $i$'s total (`global') network influence\\ 
$w_i^l$ & $i$'s influence over the top $M$ agents in $i$'s ego-network \\ 
\bottomrule
\end{tabular}
\label{tab:symbol_def}
\end{table*}

\section{Model Definition}

This simulation study performs an examination of the effect of Confederate agent selection and stance perturbation strategies on the overall stance in the network. In \autoref{tab:terminology_def} and \autoref{tab:symbol_def}, we define some of the terminology and symbols used in this study, including the scale-free influence network. At each timestep $t$, the stance vector is updated, followed by the influence matrix, using two interdependent recurrence relations. 

Following Friedkin's model \cite{friedkin}, the stance update equation represents the exogenous effects of the influence network; an agent's stance is incrementally nudged towards the average stance of those who have influence over them. The model relates influence matrix $W$ to stances $y$ at time $t$ as per \autoref{eq:stance_update}. Note, the diagonal susceptibility matrix A is scaled by a stance learning rate $\alpha = 0.001$.

\setcounter{equation}{0}
\begin{equation}
\label{eq:stance_update}
y(t) = AW(t)y(t-1) + (I-A)y(1)
\end{equation}

\autoref{eq:influence_update} describes the endogenous update rule for influence matrix $W$. Based on the Hopfield model \cite{hopfield_sim}, the influence update equation makes this a co-evolutionary model. We introduce an additional reverse relation between $W$ \& $y$ based on the homophily process, to enable influence to co-evolve with stance. We also introduces the influence update rate, $\lambda = 0.01$, also known as the rate of structural learning \cite{hopfield_sim}.


\begin{equation}
\label{eq:influence_update}
W(t+1) = \lambda y_{t} y^{\intercal}_{t} + (1-\lambda)W(t)
\end{equation}

The relation between the influence and stance update rates, $\lambda$ and $\alpha$, is crucial. Since the model dynamics are extremely sensitive to these update rates, it is also important to choose $\lambda$ and $\alpha$ such that performance comparisons between various perturbation strategies can be made. That is, when the simulation converges, it has reached a stable polarised state where there is not an absolute majority. We find that the proposed values achieve this, as demonstrated in figure \ref{fig:stance_long}. In a stable polarised state we can compare strategies based on how many agents change stances, as approximated by the average stance of non-Confederate agents.

Concretely, the case for $\alpha > \lambda$ is trivial, as the simulation converges quickly as defined by the influence network topology. Here agents adjust their stance to match their ego networks faster than they can update their influence weights based on the homophily principle of the influence update rule. Our experiments use $\lambda >> \alpha$ so that agents adjust their influence based on homophily first and foremost. The result is a changing network structure that introduces a tradeoff for Confederates between maintaining influence and perturbing stance. 


\subsection{Influence-Stance Tradeoff}
Adopting a radical position risks triggering resistance and skepticism, potentially leading to a loss of trust and influence. Operatives may establish trust by posing as like-minded individuals, sharing relatable experiences, and providing seemingly credible information. Therein, they face a dilemma; they may adopt a less extreme stance that aligns with users' existing beliefs to potentially increase their influence, or they attempt to change the narrative with extremist content and risk triggering resistance in return.

\begin{figure}[!h]
\centering
\includegraphics[width=0.8\textwidth]{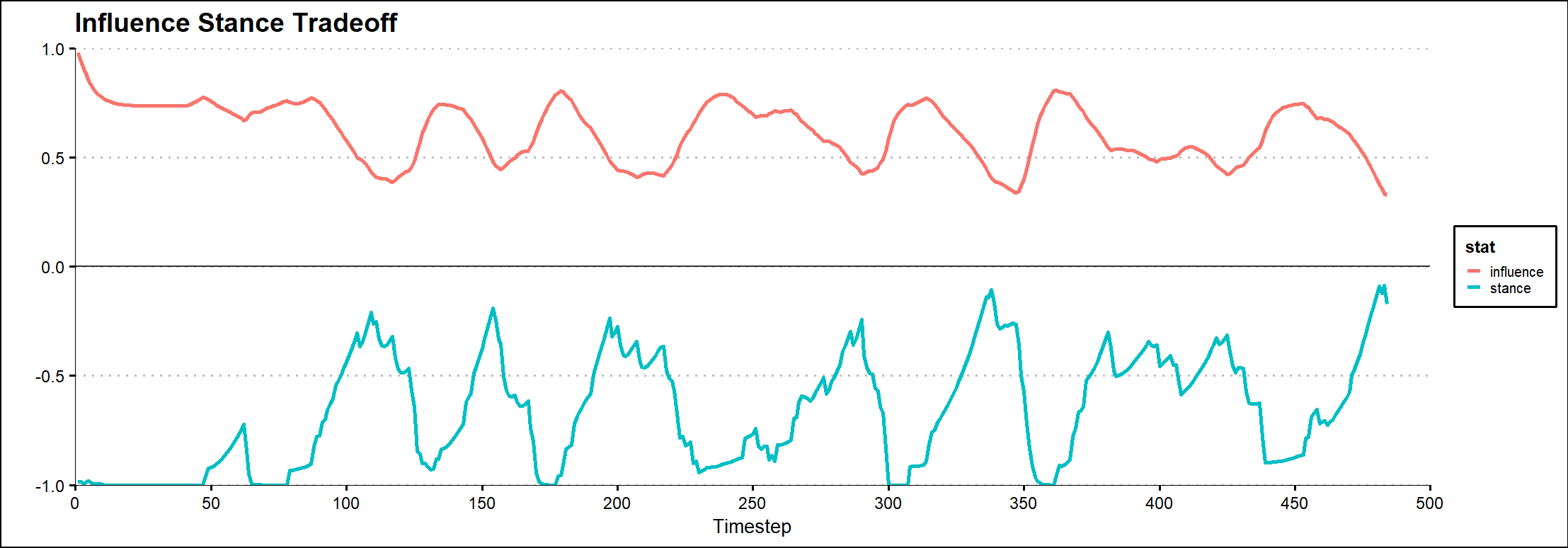}
\caption{In an 80-node network, a single Confederate struggles to perturb consensus while maintaining influence using the conversion perturbation strategy. The Confederate maintains a -1 stance until its influence begins to drop at timestep 90. It then raises its stance until its influence rebuilds. This repeats at timestep 130, the beginning of a distinctly cyclic pattern.}
\label{fig:influence_stance_tradeoff}
\end{figure}

The homophily principle in our model encapsulates this tradeoff ($\lambda y_{t} y^{\intercal}_{t}$ in \autoref{eq:influence_update}); as Confederate stances diverge from the average stance of the network, their influence decreases. 
\autoref{fig:influence_stance_tradeoff} illustrates this point with a single Confederate trying to perturb an 80-node network using the conversion perturbation strategy. 

\subsection{Confederate Selection}
With this trade-off in mind, we propose three strategies for determining Confederates.

\textbf{Maximum influence}: we select the most influential agents according to weighted out-degree in the influence matrix at time $t = 0$; $W_{max} = argmax_j \Sigma^N_i W(0)_{ij}$.

\textbf{Minimum susceptibility}: we select the least susceptible agents, skewing the remaining agents towards being more susceptible; $A_{min} = argmin_j A_{jj}$

\textbf{Random selection (control)}: Confederates are selected uniformly at random.

\subsection{Perturbation Strategy}
Confederates perturb the social network, causing the average stance of the agents to change. We define three strategies for the perturbation of a scale-free network: the conservative strategy, the conversion strategy, and the cascade strategy.

The conservative approach (\autoref{eq:conservative}) is to perturb stance if the Confederate's influence is above some threshold $\theta$. Whenever influence drops below $\theta$, we set the stance to $\mu_{y}$; the average over $N$ non-Confederate stances at time t: $\mu_{y} = \Sigma^N_i y(i, t) / N$.

\begin{equation}
\label{eq:conservative}
\!
\begin{aligned}[t]
    y(i, t) = 
     \begin{cases} 
          \mu_{y} & \Sigma^N_j w(j,i) \leq \theta \\
          -1 & \Sigma^N_j w(j,i) > \theta \\
       \end{cases}
\end{aligned}
\end{equation}

The conversion strategy (\autoref{eq:conversion}) is based on ``nudging" the network towards the desired stance (-1). It is therefore a continuous function where we scale the magnitude of the perturbation by the current level of influence as in figure \ref{fig:influence_stance_tradeoff}. When influence is low, the stance perturbation is less aggressive but when influence is high, the perturbation results in a more extreme stance (closer to -1). Here, $w^g_i$ is the global influence factor of Confederate $i$ on all $N$ non-Confederate agents (\autoref{eq:conversion_wg}). We normalize using the influence of $k$, the agent at timestep t with the maximum global influence.

\begin{equation}
\label{eq:conversion}
y(i, t) = \mu^g_{y} + w^g_i * (-1-\mu^g_{y})
\end{equation}

\begin{equation}
\label{eq:conversion_wg}
w^g_i = \Sigma^N_j w(j,i) / \max_{k'} \Sigma^N_j w(j,k')
\end{equation}


The cascade strategy (\autoref{eq:cascade}) is derived from the conversion strategy (\autoref{eq:conversion}). 
The key difference for cascade is that we take a local approach when calculating Confederate influence, looking only at each Confederate's ego network. We rank all agents by the influence the Confederate has over them and sum the influence weights from the top $M$ most influenced agents (\autoref{eq:cascade_wg}) where $M = N/10$. This more targeted approach is named after the well known cascade effect which highlights the role of network structure in information diffusion.

\begin{equation}
\label{eq:cascade}
y(i, t) = \mu^l_{y} + w^l_i * (-1-\mu^l_{y})
\end{equation}

\begin{equation}
\label{eq:cascade_wg}
w^l_i = \Sigma^M_j w(j,i) / \max_{k'} \Sigma^M_j w(j,k')
\end{equation}

\section{Methodology}

Our simulation experiment involves perturbations by a group of Confederates across scale-free networks. The ABM is implemented in the Construct framework \cite{construct, construct_1, construct_2}. This framework reads in a scale-free network and simulates the SI model as defined in Equations \ref{eq:stance_update} \& \ref{eq:influence_update} until convergence. Results are averaged over five replicates.

To begin our experiments, we construct a series of generalized scale-free networks. A scale-free network is a network where the node degrees follow a power law distribution \cite{broido2019scale}. This leads to a characteristic hub structure that mimics a social network setting, where there are some nodes that have many connections, while others are more isolated. These networks are constructed using the preferential attachment model, with five replicates per network size. Experiments on alternative network constructions such as Small World, Erdos-Renyi, and Core Periphery, are beyond the scope of this study.

We construct networks with $N$ agents, ranging from 10 to 150. We provide two attributes to each agent: stance and susceptibility. Stance, $y$, is initially 1, reflecting a state of consensus. Susceptibility, $s$ is a random variable drawn from a Normal distribution; $s\sim N(0.1,0.1)$. Next we choose a Confederate selection strategy, the percentage of agents to choose as Confederates, and a Perturbation strategy which defines the evolution of each Confederates stance. Confederate agent susceptibility is set to zero. 

Finally, we run the simulation until convergence; that is, the point where the mean stance change of all $N$ non-Confederate agents over the previous 30 timesteps is less than 0.001. At that point, the simulation is terminated. 

To measure the optimality of Confederate strategies, we calculate $\hat{\mu}_y$, the mean stance of non-Confederate agents at convergence. As such lower is better, indicating the success of Confederate perturbations in driving network stance towards -1.


\subsection{Virtual Experiments Setup}
Our virtual experiment setup, detailing the independent, dependent and control variables are summarized in \autoref{tab:virutal_expt_table}.
\setlength{\tabcolsep}{2pt}
\begin{table*}[!ht]
\centering
\caption{Definitions of Terminology and Symbols used in this study}
\begin{tabular}{lcccc}
\toprule
\textbf{Variable} & \textbf{type} & \textbf{Range} & \textbf{Value} & \textbf{Number} \\ 
\midrule
Number of agents, $N$ & Independent & [10, 150] & 10, 20, 30... 150 & 14 \\ 
Percent Confederates & Independent & [0, 100] & 5, 10, 15,..., 40 & 8 \\ 
Perturbation strategy & Independent & & Equations \ref{eq:conservative}, \ref{eq:conversion}, and \ref{eq:cascade}  & 3 \\
Agent selection strategy & Independent & & $W_{max}, A_{min},$ Random & 3 \\
Mean stance at convergence, $\hat{\mu}_y$ & Dependent & [-1.0, 1.0] & [-1.0, 1.0] & $\mathbb{R}$ \\ 
Convergence timestep & Dependent & $\mathbb{N}$& [50, 200] & $\mathbb{N}$ \\
Initial stance, $y$ & Control & [-1.0, 1.0] & -1, 1 & 2 \\ 
Susceptibility, $s$& Control & [0, 1.0] & $s\sim N(0.1,0.1)$ & $\mathbb{N}$ \\ 
Influence update rate, $\lambda$ & Control &[0, 1.0] & 0.01 & 1 \\
Stance update rate, $\alpha$ & Control &[0, 1.0] & 0.001 & 1 \\
\bottomrule
\end{tabular}
\label{tab:virutal_expt_table}
\end{table*}

\section{Results and Discussion}
We detail four key results; convergence to a polarized state, optimal Perturbation strategies, optimal Confederate selection strategies, and the observation of tipping points.

\begin{figure}[!h]
\centering
\includegraphics[width=0.8\textwidth]{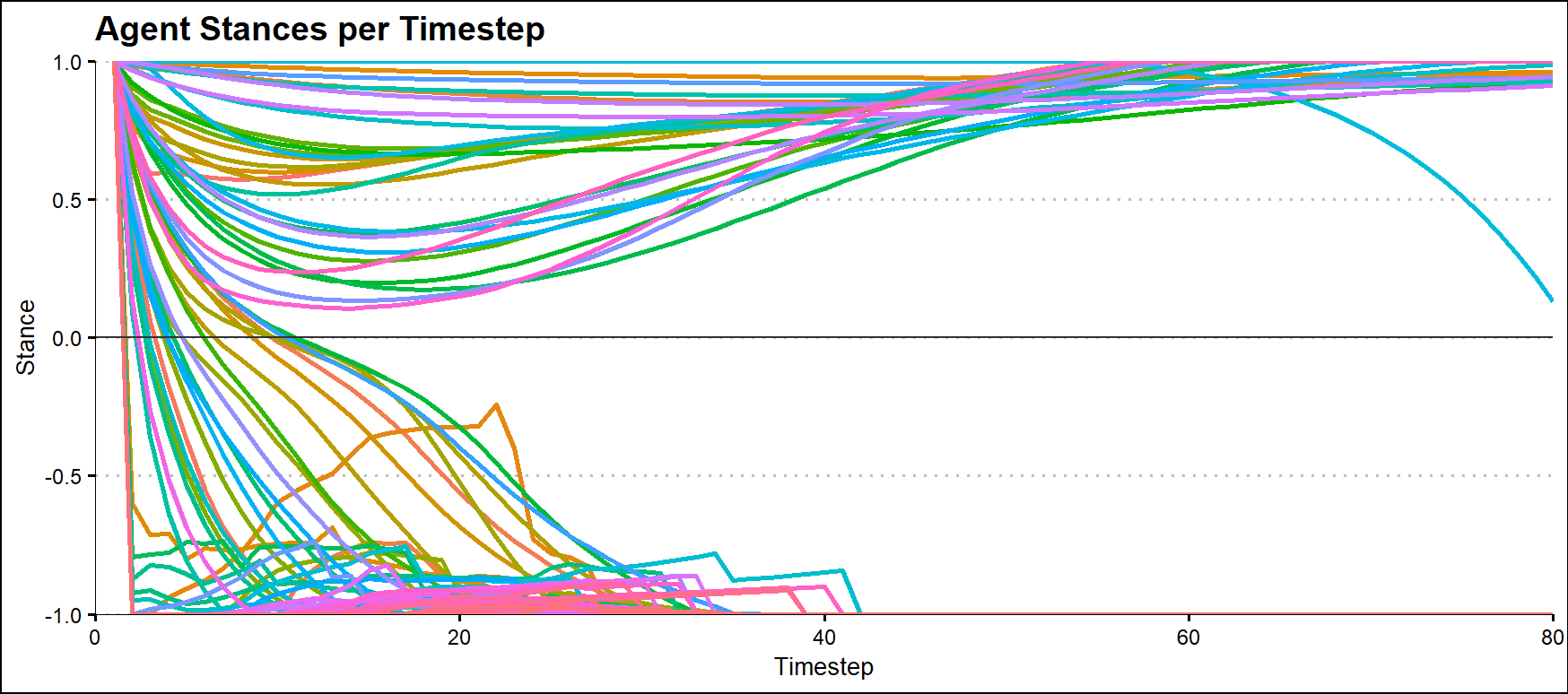}
\caption{Simulation of the change in agent stances for an 80-node network. Each line represents an agent's stance over time. Stances eventually converge into one of two extremes, 1 and -1.}
\label{fig:stance_long}
\end{figure}

\paragraph{\textbf{Convergence to a state of stable polarization}}
We first observe that stances in a network do eventually come to convergence. \autoref{fig:stance_long} illustrates the change in agent stances across time for an 80-node network, where each line represents an agent. The stances in the simulation eventually converge into one of the two extremes, 1.0 and -1.0. 

\begin{figure}[!h]
  \centering
  \begin{minipage}{0.48\textwidth}
    \includegraphics[width=\textwidth]{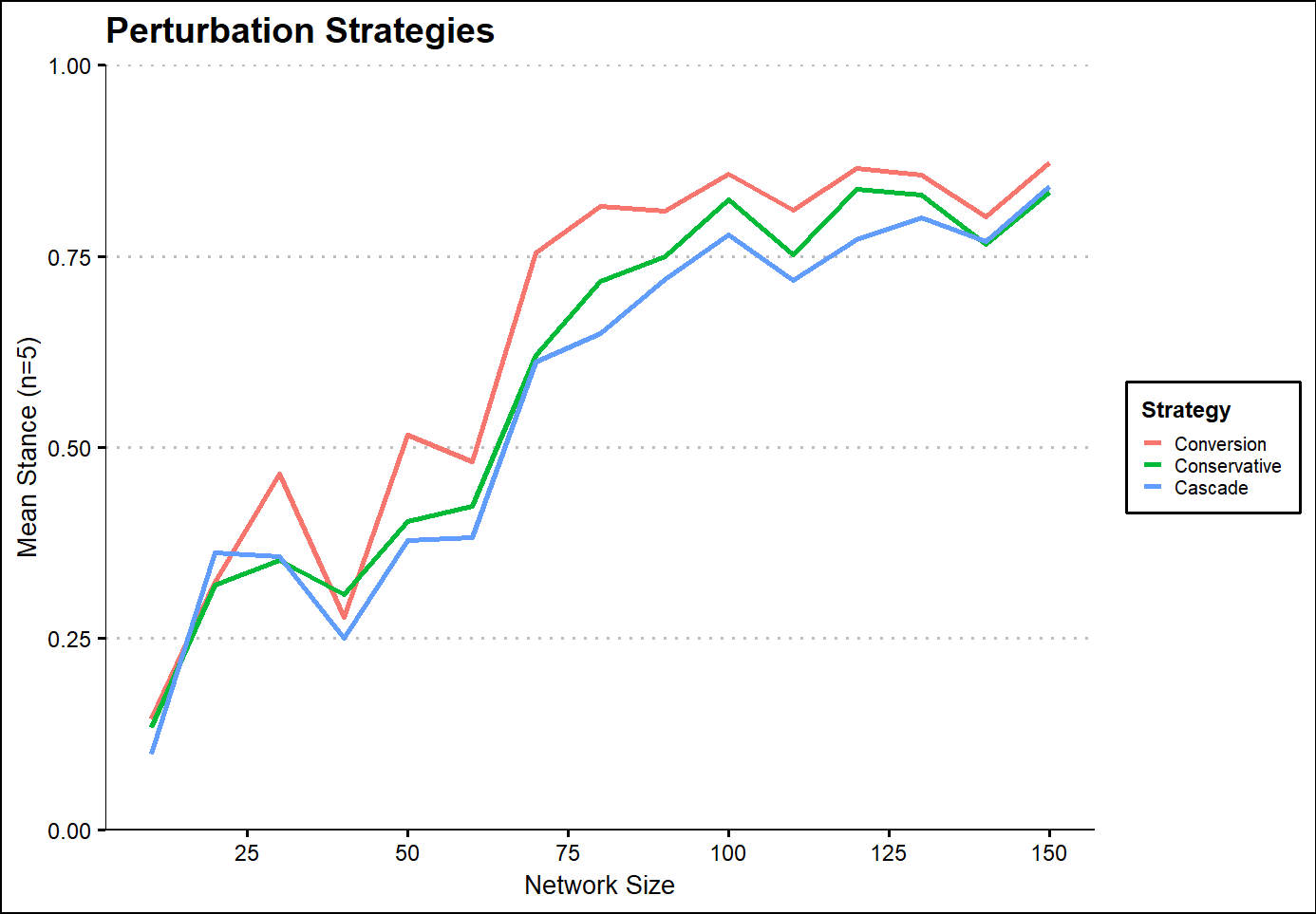}
    \caption{Comparison of the perturbation strategies, lower is better. The cascade strategy is optimal.}
    \label{fig:optimal_conferates}
  \end{minipage}
      \hfill
  \begin{minipage}{0.48\textwidth}
    \includegraphics[width=\textwidth]{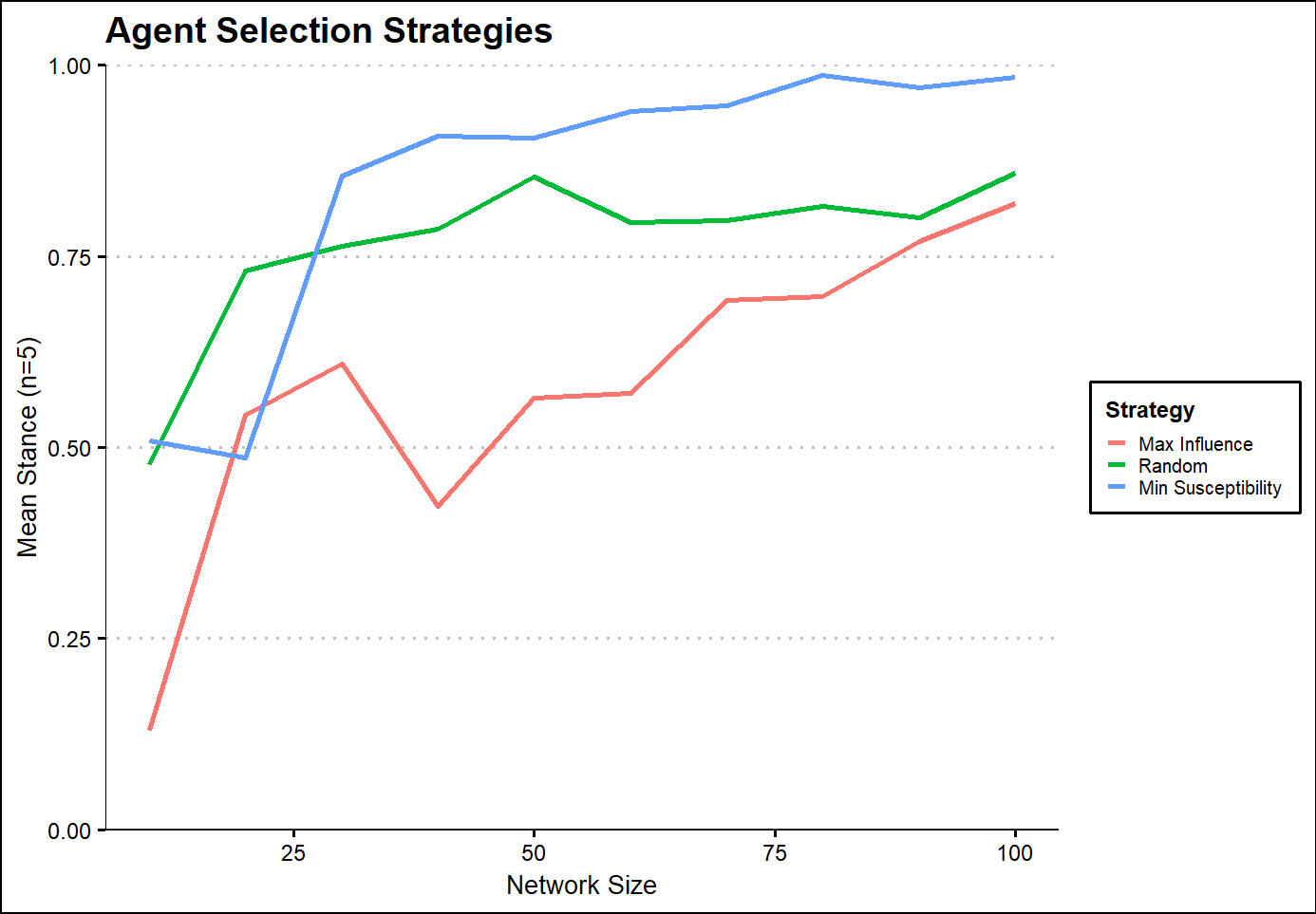}
    \caption{Comparison of the agent selection strategies, lower is better. Influential Confederates are optimal.}
    \label{fig:influential_agents}
    \end{minipage}
\end{figure}

\paragraph{\textbf{Cascade is optimal}}
In examining the perturbation strategies against network size and means stances, we observe that the optimal Perturbation strategy is the cascade strategy. That is, an optimal strategy for the changing of stances involves the nudging of agents that are in a Confederate agent's direct neighborhood. This is illustrated in \autoref{fig:optimal_conferates}, where the line of mean stance for the cascade strategy is the lowest (best). 
We also note that the conversion strategy fares the worst out of the three strategies, indicating that global nudging strategies are not optimal in general and require precise targeting.

\paragraph{\textbf{Influential Confederates are best}}
Out of the three Confederate selection strategies, influential agents are the best Confederates. This is observed in \autoref{fig:influential_agents}, where Confederate agent selection strategies are plotted against each other. We also note a clear trend of diminishing returns where Confederate strategies are less successful as the network size increases. This hints at the underlying resilience of scale free networks. 

\begin{figure}[!htbp]
\centering
\includegraphics[width=0.8\textwidth]{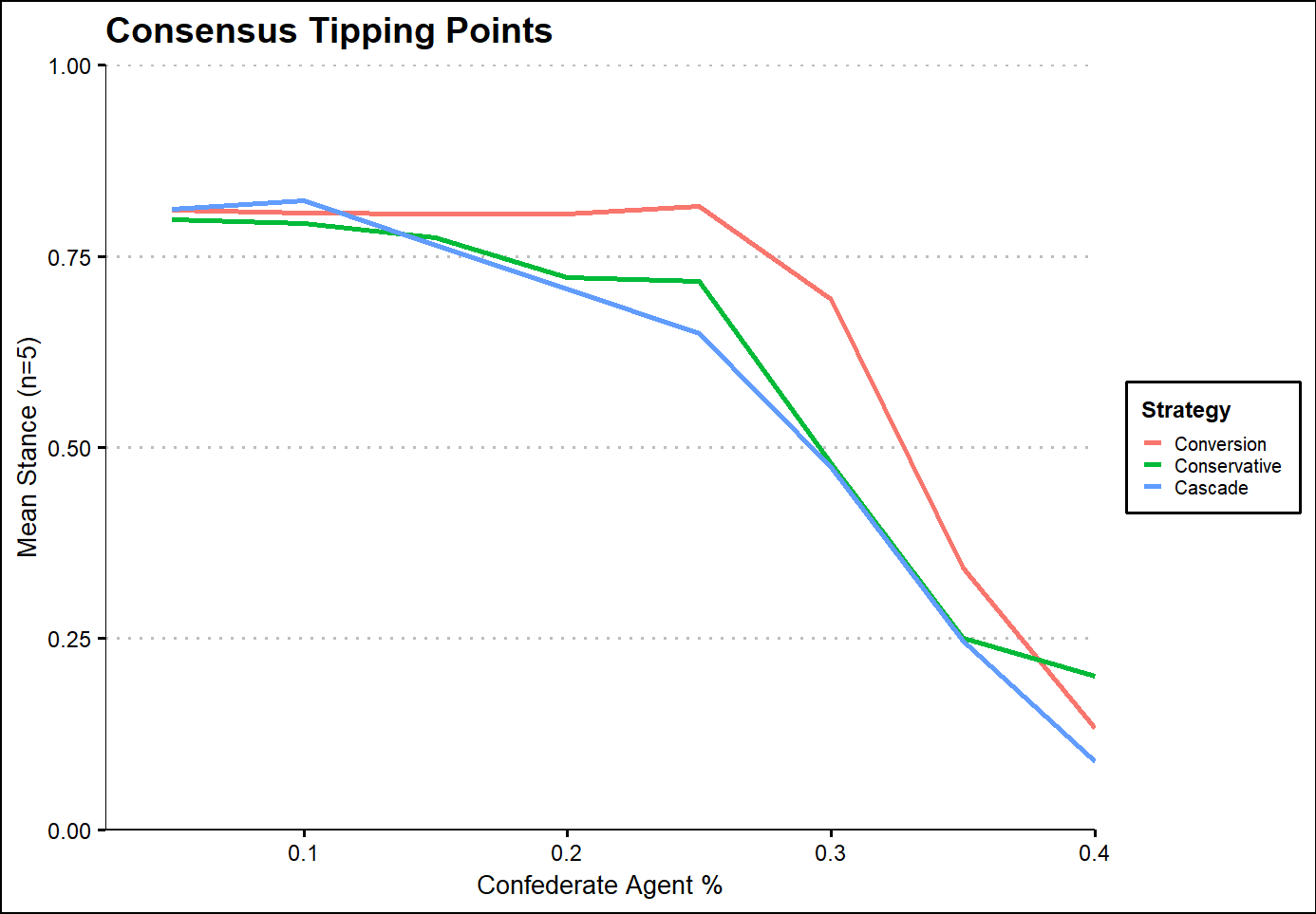}
\caption{With $>$20-25\% of network agent as Confederates, mean stance shifts rapidly. Results are averaged over five 80-node networks, using the maximum influence selection strategy.}
\label{fig:minority_stance}
\end{figure}

\paragraph{\textbf{Minority stance tipping points exist}}
We observe that Confederate agents can create a tipping point wherein the network converges to the stance of the minority Confederate agents. \autoref{fig:minority_stance} shows tipping point ranges between 20-25\% for the three different Perturbation strategies applied, mirroring empirical evidence from Centola et. al. \cite{tipping_points}.  
The cascade strategy requires the least Confederates to cause the tip, while the conversion strategy requires the most Confederates to tip the overall stance. This mirrors the findings for the stance perturbation experiment (\autoref{fig:optimal_conferates}).




\section{Validation}
Our results match stylized facts that can be used to determine known phenomena. States of consensus are reached when users achieve social conformity with each other \cite{disinfo_sim, construct_1, construct_2}, which is observed through stance convergence (\autoref{fig:stance_long}) and tipping points (\autoref{fig:minority_stance}) within our work. This is also observed within the principles of homophily, giving rise to the adage ``birds of a feather flock together".

We find that influential people are best positioned to affect other people's stances (\autoref{fig:influential_agents}). Studies on targeted attacks on scale-free networks have shown that higher degree nodes have larger effects on the overall network and our experimental results provide the same observation \cite{manipulative_bots, jalili2017information}. Similarly, the optimal perturbation strategy in our model is to begin by effecting a change towards people that are close to you, i.e. your ego-network (\autoref{fig:optimal_conferates}). This is in-line with the theory of information cascades \cite{jalili2017information}.

Finally, the tipping points observed in our experimental runs are similar to previous research in the literature \cite{tipping_points}; 20-25\% of the network are required as Confederates to change the stance of the network. These results lend our model theoretic validity and face validity.

\section{Limitations and Future Work}
Validation is challenging; it is clearly unethical to precipitate consensus change in a real social network. A lack of real-world data further complicates the matter, as observations of stance changes in real-world networks are extremely rare. Ng et. al. \cite{stance_flip} find that only 1\% of Twitter users changed their stances towards the Coronavirus vaccine over a one-year period. Most Twitter users maintained the same stance. An area of future research is to identify and process real event data with stance and opinion annotation change across time and adjust our formulation to the observed data. 

Despite these limitations, it is important to understand the circumstances under which agents in a social situation will change their opinion, which has implications for persuasion and disinformation research. Differences in social networking platforms may play a role; where account histories are easily accessible, for instance, modeling suspicion as a function of stance volatility would counteract certain Confederate strategies. To account for this, \autoref{eq:influence_update} could be modified to incorporate a regularization parameter that penalizes highly volatile agents. Additionally, experimenting with other models for the influence network such as Small World, Erdos-Renyi, and Core Periphery, may explain how network structure affects the success of the Cascade strategy. 

Given the emergence of stable polarized states within this model, perhaps the most natural and pertinent avenue for future work is the recovery from such a state. A body of theory, and in particular Krackhardt's notion of simmelian ties, supports such an effort \cite{ties_that_torture}. The core building block of our model is that of the dyadic tie. Extending this model to capture triadic relationships would enable the interrogation of the role non-prejudiced third parties play in recovering from the polarized state.



\section{Conclusion}
In this work, we formulate and test a co-evolutionary social influence model to simulate intentional stance perturbation in scale-free networks. We define three strategies for selecting a set of Confederates that will perturb the network: maximum influence, minimum susceptibility, and a random selection for a control experiment. We design three Perturbation strategies that Confederates use to probe the network: conservative, conversion, and cascade strategies. Our results show that influential agents are the best choice of Confederate, that the optimal Perturbation strategy involves the targeted nudging of local ego networks and that there exists a range of tipping points for group consensus. We hope that this simulation sheds light on the effectiveness of intentional stance change within a social network and the manner in which successful social influence operations are run.

\section{Acknowledgements}
This work was supported in part by the Office of Naval Research grant (N000141812106) and the Knight Foundation. Additional support was provided by the Center for Computational Analysis of Social and Organizational Systems (CASOS) at Carnegie Mellon University. The views and conclusions in this document are those of the authors and should not be interpreted as representing the official policies, either expressed or implied, of the Knight Foundation, Office of Naval Research, or the U.S. Government.

\footnotesize
\bibliographystyle{splncs04}
\bibliography{stance_paper.bib}

\end{document}